\documentclass[3p,times]{elsarticle}

\usepackage{ecrc}
\usepackage{xspace}
\usepackage{subfigure}
\usepackage[capitalize]{cleveref}
\usepackage{microtype}

\volume{00}

\firstpage{1}

\journalname{Nuclear Physics A}

\runauth{Benjamin Bannier (for the PHENIX collaboration)}

\jid{npa}

\jnltitlelogo{Nuclear Physics A}

\usepackage{amssymb}

\usepackage{lineno}

\biboptions{square,comma,numbers,sort&compress}

\usepackage[figuresright]{rotating}

\newcommand{\pt}{\ensuremath{{p_T}}\xspace}
\newcommand{\Rg}{\ensuremath{{R_\gamma}}\xspace}
\newcommand{\ef}{\ensuremath{{\langle\varepsilon f\rangle}}\xspace}
\newcommand{\piz}{\ensuremath{{\pi^0}}\xspace}
\newcommand{\D}{\ensuremath{\text{d}}\xspace}
\newcommand{\text}[1]{\ensuremath{\mathrm{#1}}}
\begin{document}

\begin{frontmatter}



\dochead{}

\title{Measurements of direct photons in Au+Au collisions with PHENIX}


\author{Benjamin Bannier (for the PHENIX collaboration)}

\address{Department of Physics and Astronomy, Stony Brook University, Stony Brook, NY 11794, USA}

\begin{abstract}
  The PHENIX experiment has published direct photon yields and elliptic flow
  coefficients $v_2$ from Au+Au collisions at RHIC
  energies~\cite{yield_PHENIX,v2_PHENIX}. These results have sparked much
  theoretical discussion. The measured yields and flow parameters are difficult
  to reconcile in current model calculations of thermal radiation based on
  hydrodynamic time evolution of the collision volume. %
  Our latest analyses which use high statistics data from the 2007 and 2010
  runs allow the determination of direct photon yields with finer granularity in
  centrality and photon momentum and down to \pt as low as $0.4\,\text{GeV}/c$.
  We will summarize the current status and present new results from PHENIX.
\end{abstract}

\begin{keyword}
Direct photons



\end{keyword}

\end{frontmatter}


\section{Introduction}
\label{sec:introduction}

Direct photons which are photons not produced in late hadron decays have been
considered an excellent tool to characterize the strongly interacting states of
matter produced in heavy-ion collisions for a long time. Photons are produced
during all stages of the evolution of the colliding system and transport
information about their production environment to the detector virtually
undistorted due to their very small interaction cross section with the strongly
interacting medium.

While the large yield of soft direct photons with $1\,\text{GeV}/c \lesssim \pt
\lesssim 3\,\text{GeV}/c$ seen in central Au+Au collisions at
RHIC~\cite{yield_PHENIX} hints at an enhanced production of \emph{thermal}
direct photons from an early, hotter medium, the observation of strong
elliptical flow $v_2$ of direct photons under similar
conditions~\cite{v2_PHENIX} where direct photons showed a $v_2$ \emph{en par}
with that of hadrons seemed an indication of very late production.
Similar observations where made at LHC energies in Pb+Pb
collisions~\cite{yield_ALICE,v2_ALICE}.
Accommodating both a large direct photon yield and flow has been a challenge
for models.
To provide further constraints on mechanisms of direct photon production PHENIX
has measured the centrality-dependence of the direct photon yield and extended the \pt-coverage of the measurement to lower momenta $\pt>0.4\,\text{GeV}/c$.

\section{Method}
\label{sec:method}

Since measurements of soft photons in electromagnetic calorimeters are
notoriously difficult due to the large hadronic backgrounds in heavy-ion
collisions we reconstruct real photons from electron-positron pairs produced in
external conversions of photons in the detector material~\cite{rich_proc} which
allows us to take advantage of good electron reconstruction
capabilities with PHENIX down to $\pt=0.2\,\text{GeV}/c$. With this method we are able to
reconstruct photons down to twice the minimal momentum for electrons, i.e.\ 
$\pt=0.4\,\text{GeV}/c$.
To
minimize dependence on exact knowledge of e.g.\ the material budget of the
detector we measure the direct photon yield indirectly via a double ratio

\begin{equation}
  \Rg = \frac{Y^\gamma_\text{incl}}{Y^\gamma_\text{hadron}}
      = \frac
      {\ef \left(\frac{N^\gamma_\text{incl}}{N^\gamma_\piz}\right)_\text{data}}
      {\left(\frac{Y^\gamma_\text{hadron}}{Y^\gamma_\piz}\right)_\text{sim}}
\label{eq:rg}
\end{equation}

where $Y^\gamma_\text{incl}$ and $Y^\gamma_\text{hadron/\piz}$ denote yields of
photons \emph{produced} inclusively or in hadron/\piz decays, and
$N^\gamma_\text{incl}$ and $N^\gamma_\piz$ are the \emph{observed} inclusive
photon yield and the \emph{observed} yield of photons from \piz decays which.
We directly measure the \piz decay photon contribution to the inclusive photon
yield (which represents $\sim 80\%$ of it) by tagging.
The
factor \ef corrects for the finite efficiency with which photons from \piz
decays can be tagged experimentally. With this definition $\Rg \geq 1$ always,
and $\Rg>1$ is an observation of direct photons.

\subsection{Inclusive photon sample $N^\gamma_\text{incl}$}

Since the PHENIX tracking system~\cite{PHENIX_tracking} uses detectors operated
$2\,\text{m}$ or further away from the interaction vertex track reconstruction
is done assuming particle tracks originating from the nominal vertex position.
In this procedure momenta of particles originating from off-vertex locations
$R>0$ are misreconstructed since the actual bending angle of the particle due
the magnetic field is smaller than when assuming travel through the full
magnetic field. Since the PHENIX magnetic field is roughly parallel to the beam axis for
measurements in the central arm detectors this leads to off-vertex particles
being reconstructed with artificially larger momenta. By the same process
electron-positron pairs from off-vertex conversions are reconstructed with
larger and larger masses the further out radially the conversion occurred.

During the 2007 and 2010 RHIC heavy-ion runs the Hadron Blind Detector
(HBD)~\cite{HBD} was installed before the tracking detectors whose backplane
components (readout boards and chips, and electronics) at radial positions
$R\approx 60\,\text{cm}$ provide spatially defined conversion locations with
$X/X_0 \approx 3\%$. Momenta of tracks were calculated assuming
production at the vertex, and assuming production from the $R\approx
60\,\text{cm}$. Electron-positron pairs from conversions in the HBD backplane
have fake invariant masses $m_\text{ee}>0$ in the standard procedure, but are
seen with $m_\text{ee}\approx 0$ under the off-vertex assumption. Conversely,
electron-positron pairs from \piz-Dalitz decays have correct mass
$m_\text{ee}\approx 0$ in the standard procedure, but a fake $m_\text{ee}>0$
assuming off-vertex production. By selecting on both masses we are able to
extract a sample of real photons with purity $>99\%$.

The observed yield of HBD backplane conversion pairs is related to the actual
photon yield by detector- and reconstruction-dependent factors,
$N^\gamma_\text{incl} = Y^\gamma_\text{incl} p_\text{conv} a_\text{ee} \varepsilon_\text{ee}$,
specifically the conversion probability in the HBD backplane $p_\text{conv}$, the
geometrical acceptance for the electron-positron pair $a_\text{ee}$ and
reconstruction efficiency of the pair $\varepsilon_\text{ee}$. We measured
$N^\gamma_\text{incl}$ as a function of the converted photon \pt.

\subsection{\piz-tagged photon sample $N^\gamma_\piz$}

To measure the yield of photons from \piz decays we use an additional photon
detected conventionally in the electromagnetic calorimeter which we pair with
the photon reconstructed in the conversion pair. For the second photon we use
very loose cuts in order to maintain maximal photon efficiency and to minimize
the effect of systematic uncertainties on the final result. After modelling the
combinatorial background in photon-photon pairs with a mixed event technique
the observed photon yield from \piz decays is determined by integrating mass
peaks in the photon-photon mass spectrum around the \piz mass. Since we
measure the \piz-tagged yield as a function of converted photon \pt, the
converted photon introduces the identical detector- and reconstruction-specific
factors as for the inclusive sample, but additionally depends on the conditional
acceptance-efficiency to reconstruct the second photon from a \piz decay in the
calorimeter given we already reconstructed the first photon in a conversion
pair, \ef. Specifically we have
$N^\gamma_\piz = Y^\gamma_\piz p_\text{conv} a_\text{ee} \varepsilon_\text{ee}
\times \ef$. Now in the upper ratio in Eq.~(\ref{eq:rg}) the detector- and
reconstruction dependent factors cancel explicitly and it only depends on the
physical yields and \ef.

\subsection{The tagging efficiency correction \ef}

The tagging efficiency \ef depends on the geometrical acceptance for
photons measured in the calorimeter $f$ and their reconstruction efficiency $\varepsilon$.
Since we use only very loose cuts on the second photon $\varepsilon\approx
90\%$, while $f$ can be calculated accurately from the known detector geometry
and knowledge of the distribution of live channels in the calorimeter. With
known kinematic distributions of parent \piz and decay kinematics
the centrality-dependent factors \ef were then
calculated in a Monte Carlo simulation as functions of converted photon \pt.

\subsection{The ratio $\frac{Y^\gamma_\text{hadron}}{Y^\gamma_\piz}$}
\label{sec:cocktail_ratio}

Finally, the yield ratio between all hadron sources and exclusively from \piz
decays can be calculated in a Monte Carlo simulation independent of the
detector specifics. We include the following photon production channels in our calculation:
$\piz\rightarrow \gamma\gamma$,
$\eta\rightarrow \gamma\gamma, \pi^+\pi^-\gamma$,
$\eta^\prime\rightarrow \gamma\gamma, \pi^+\pi^-\gamma, \omega\gamma$,
$\omega\rightarrow\piz\gamma$.
The shapes of \piz \pt spectra are derived from $\pi$ data~\cite{PPG088} and
derived assuming $m_T$ scaling for the shape for other hadronic sources.
The meson/\piz ratios are constrained by data~\cite{PPG088,Adare:2012wg}.

\section{Results}
\label{sec:results}

\begin{figure}[hbt]
  \subfigure[\Rg measured in external conversion pairs in the 2007 and 2010
      data set (filled circles and squares) and from the virtual photon
      analysis~\cite{yield_PHENIX} (open circles).]{%
      \includegraphics[width=0.49\linewidth]{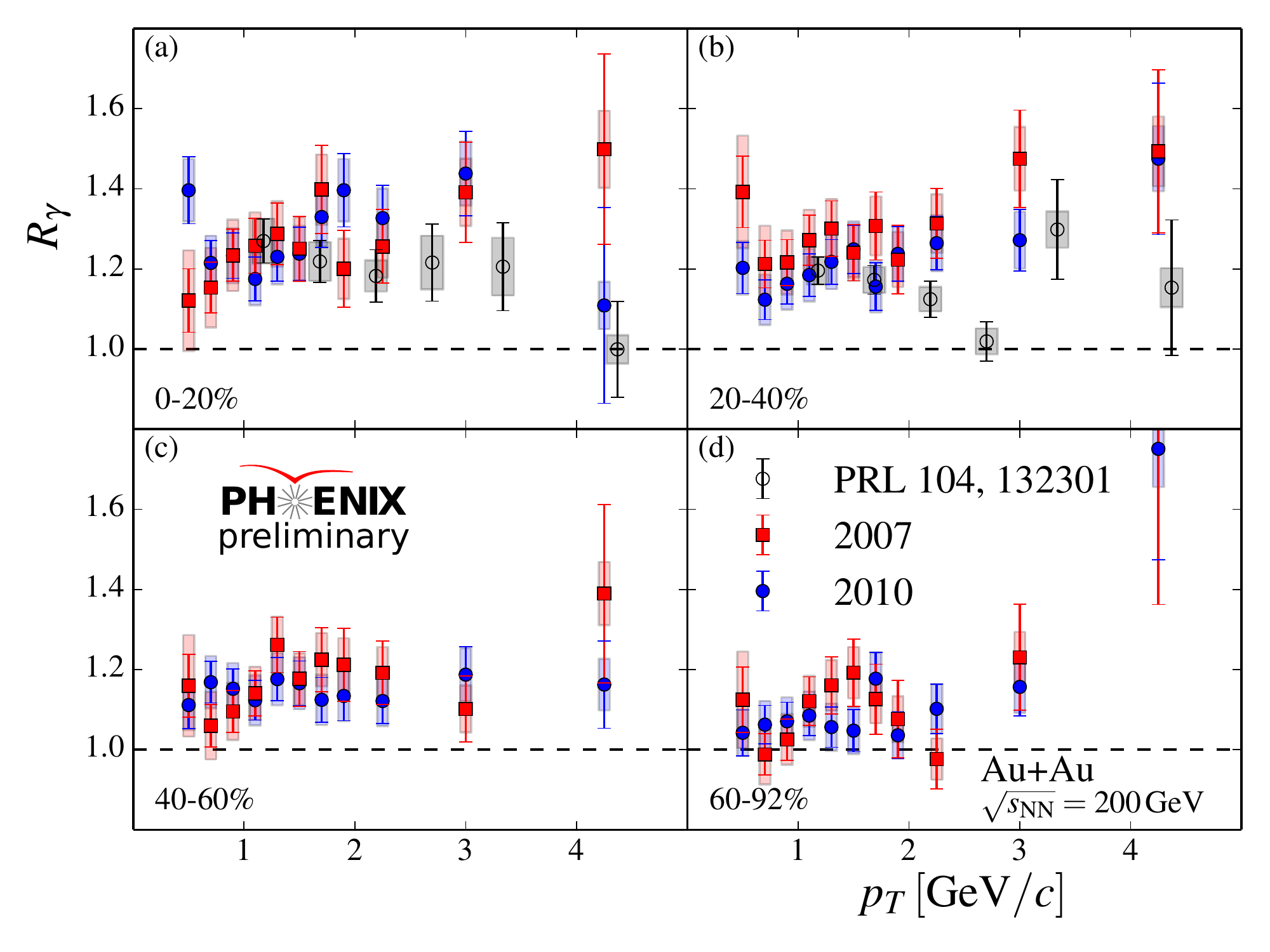}
      \label{fig:rg}}%
  \subfigure[
    The direct photon \pt spectrum measured in external conversion pairs from
    the 2007+2010 combined data set(blue circles) and from the virtual photon
    analysis~\cite{yield_PHENIX}. The green line is an $N_\text{coll}$-scaled fit to
    the $p+p$ data, see the text for details.]{%
      \includegraphics[width=0.49\linewidth]{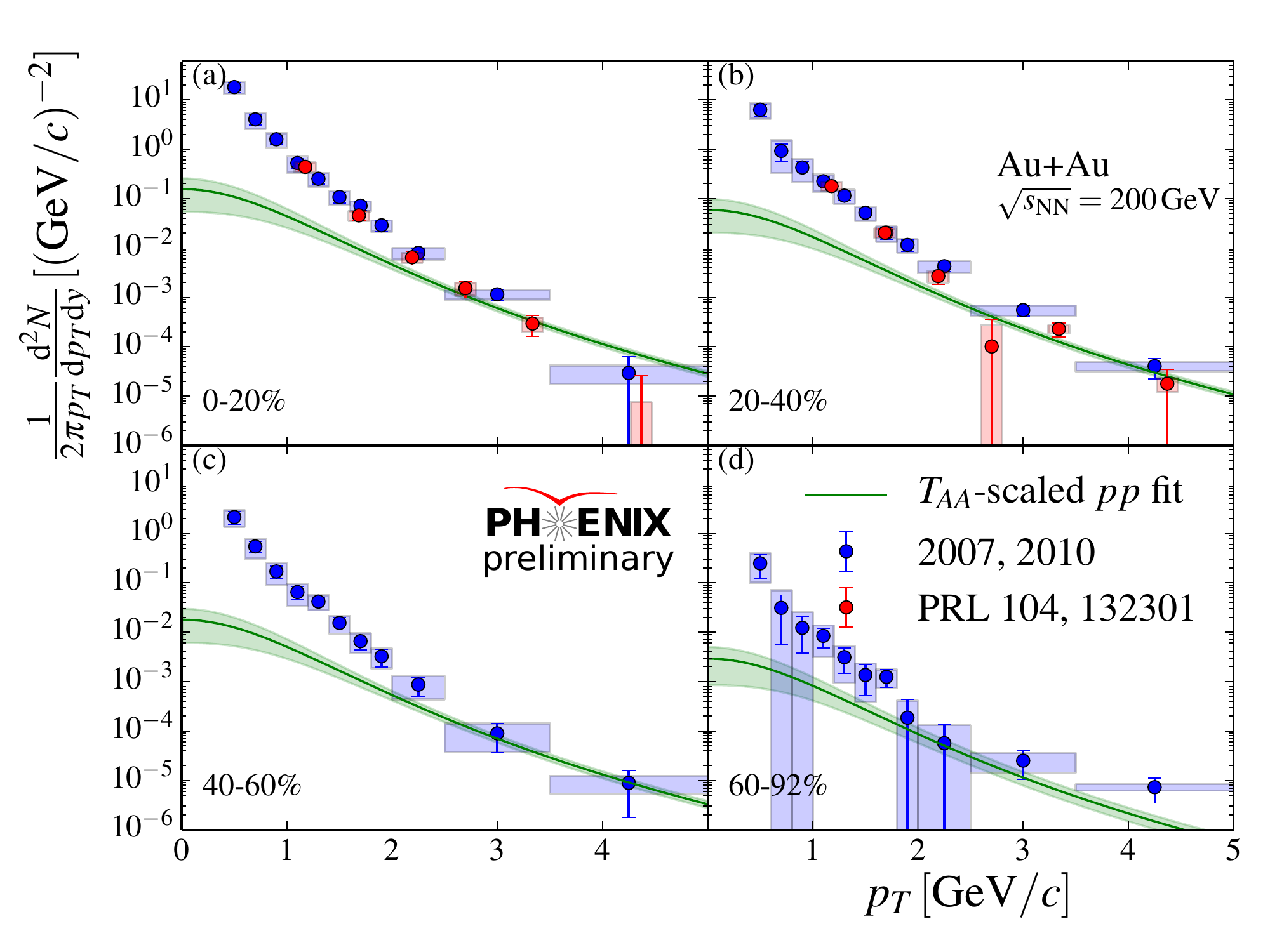}
      \label{fig:ydir}
  }
  \caption{}
\end{figure}

The \Rg measured in the 2007 and 2010 data set in centrality classes 0-20\%,
20-40\%, 40-60\% and 60-92\% are shown in Fig.~\ref{fig:rg} where for the 2007 data
set \ef was determined with full Geant simulation while for the 2010 data set
it was determined in a fast Monte Carlo simulation. We obtain consistent
results between the two measurements, and can confirm the previous virtual
photon analysis. We find that \Rg approaches unity when going to more
peripheral collisions. With the additional measurements at lower momenta we
find at best a weak dependence of \Rg on the photon momentum.
In the following we will use run-specific corrections \ef determined with the
fast Monte Carlo simulation for both the 2007 and 2010 data sets.  The combined
2007+2010 can be calculated from the weighted average of the data ratios
(numerators in Eq.~(\ref{eq:rg})) where systematic uncertainties are completely
correlated. The denominators in Eq.~(\ref{eq:rg}) do not depend on the specific
year.

\begin{figure}[hbt]
  \subfigure[The excess photon spectra in external conversion pairs from the
    2007+2010 combined data set. The dashed lines are exponential fits in the range
    $0.6\,\text{GeV}/c < \pt < 2.0\,\text{GeV}/c$]{%
    \includegraphics[width=0.49\linewidth]{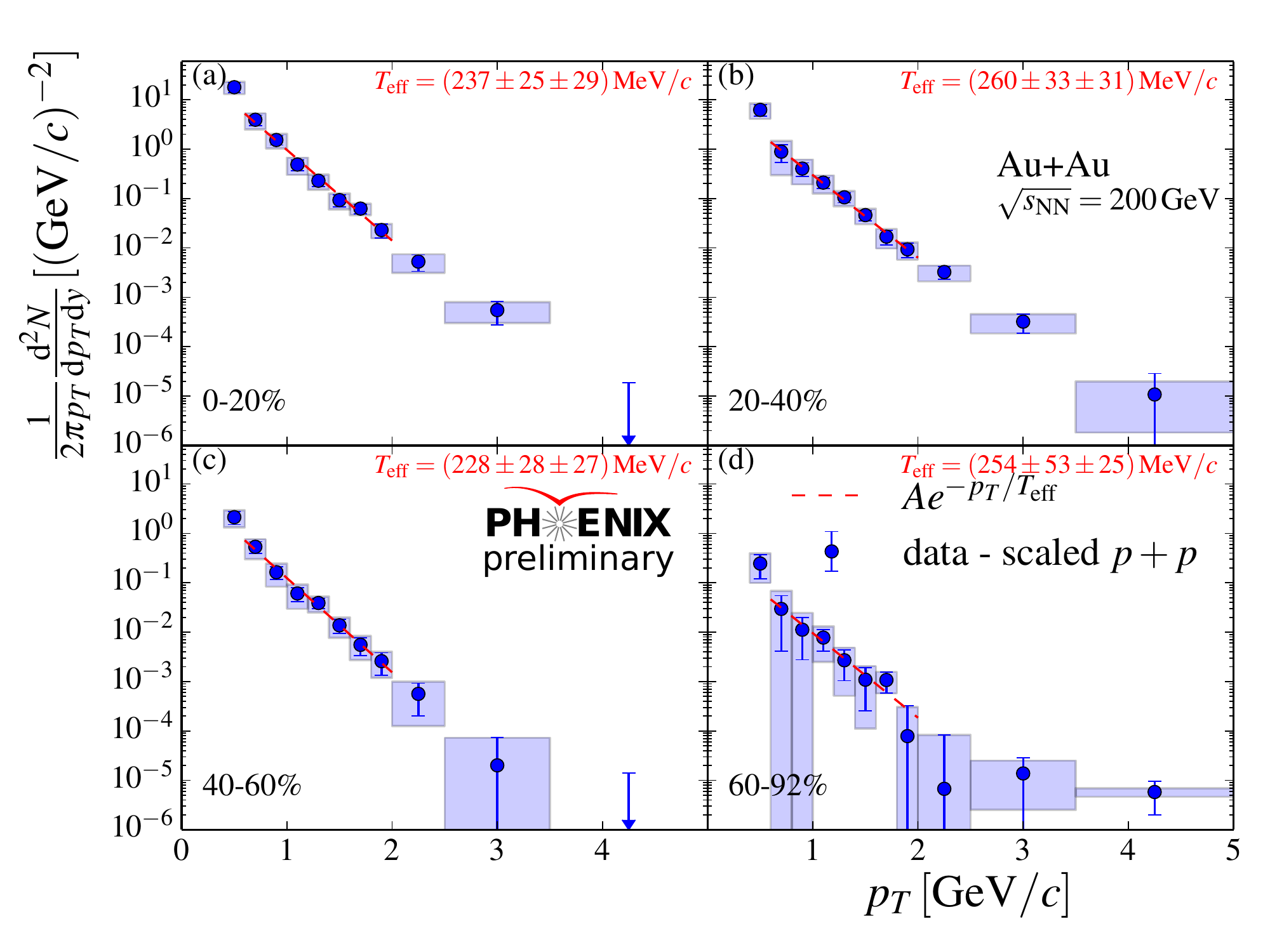}
    \label{fig:yextra}
  }%
  \subfigure[The integrated excess photon yield as a function of $N_\text{part}$ for
    different lower integration limits. The dashed lines are independent fits
    $AN_\text{part}^x$ to each set of points.]{%
    \includegraphics[width=0.49\linewidth]{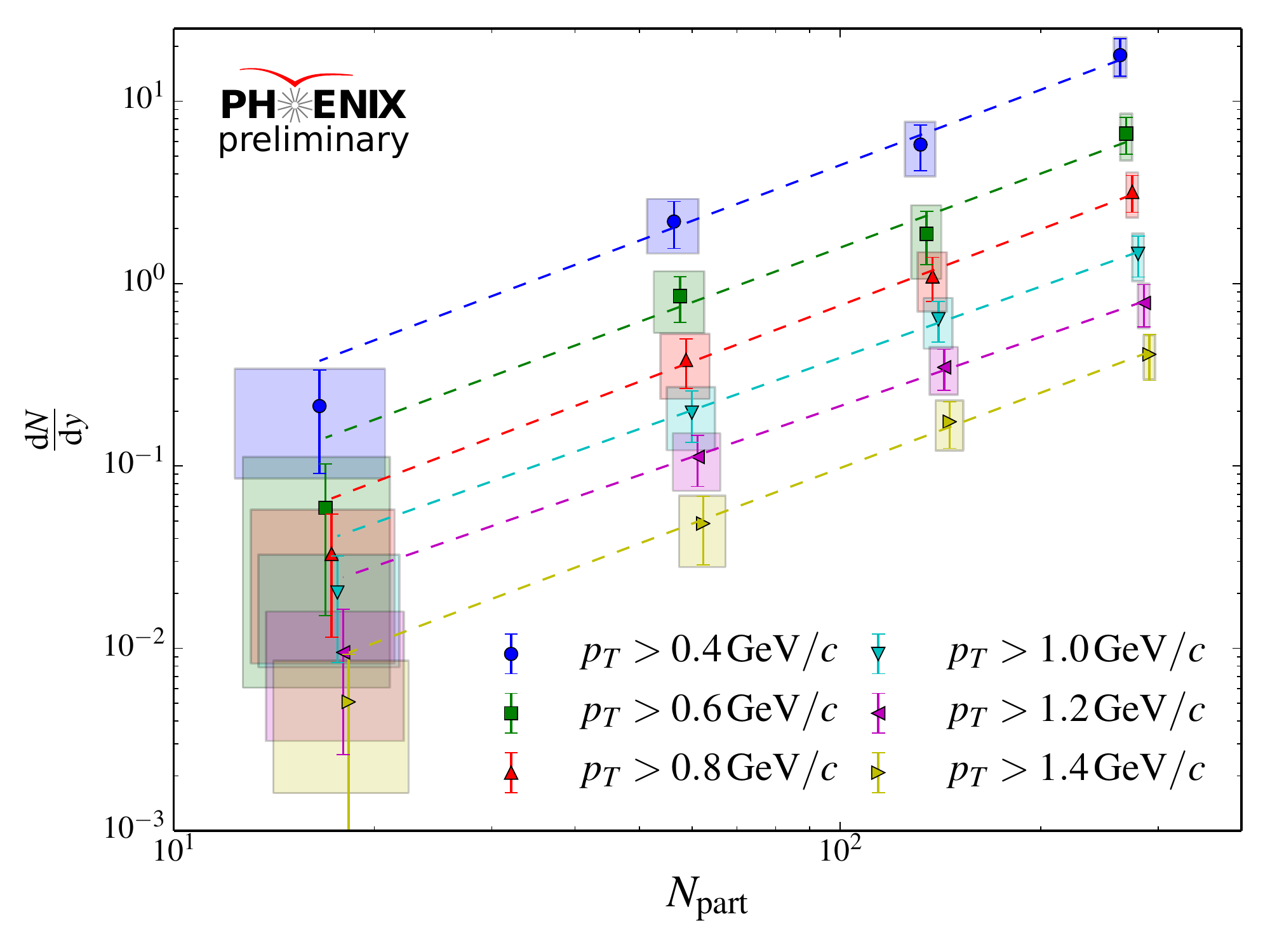}\label{fig:inty}}%
  \caption{}
\end{figure}


With the cocktail of hadronic photon sources described in
Section~\ref{sec:cocktail_ratio} we can calculate the direct photon momentum spectra
shown in Fig.~\ref{fig:ydir} from $Y^\gamma_\text{direct} = (R_\gamma - 1)
Y^\gamma_\text{hadron}$ with $Y^\gamma_\text{direct, hadron}$ the physical yields
of direct and hadron decay photons, respectively.
To isolate the purely medium-induced component of the direct photon signal we
parametrize the direct photon signal from prompt production measured in $p+p$
collisions~\cite{yield_PHENIX,Adler:2006yt,Adare:2012yt} with a modified power
law $a\left(1+\frac{\pt^2}{b}\right)^{-c}$ which we scale for each centrality
class according to the number of binary collisions, also shown in
Fig.~\ref{fig:ydir}. After subtraction of this hard component from the scaled
$p+p$ fits we arrive at the excess photon spectra shown in Fig.~\ref{fig:yextra}.
The excess spectra are clearly not exponential in the shown momentum range which might
be expected given the non-zero photon flow. To characterize the shape of the
spectra in the low-momentum range we fit exponentials $A e^{-\pt/T_\text{eff}}$
in the range $0.6\,\text{GeV}/c < \pt < 2.0\,\text{GeV}/c$ where the inverse
slope parameter $T_\text{eff}$ should be interpreted exclusively as a shape parameter
for this discussion since due to direct photon flow extractions of the medium
temperature from the spectra are not obvious~\cite{Shen:2013vja}.
We find remarkably similar values
for the inverse slope
across centralities within the given uncertainties while the yield varies over more than 2 orders of
magnitude. The inverse slopes extracted here are overall compatible with the
earlier PHENIX measurement from virtual photons~\cite{yield_PHENIX} which where
extracted in a slightly higher momentum range. We find that the direct photon
spectra at RHIC energies are softer than at LHC where ALICE has extracted
$T_\text{LHC} = (304\pm 51^{\text{syst+stat}})\,\text{MeV}$~\cite{yield_ALICE}.

To characterize the centrality-dependence of the excess photon spectra we
calculate integrated yields,

\[
  \frac{\D N}{\D y} = 2\pi \int_{p_{T,\text{min}}}^{5\,\text{GeV}/c} \D\pt \pt \left(\frac{1}{2\pi\pt}\frac{\D^2N}{\D\pt\D y}\right).
\]

Since the excess photon spectra are steeply falling functions of photon \pt the
integrated yields will depend mostly on the lowest few measurements in the
spectra. To not prefer one momentum range over another we calculate integrated
yields for all possible lower integration limits in the range
$0.4\,\text{GeV}/c < \pt < 1.4\,\text{GeV}/c$ while leaving the upper limit
fixed at the end of our measured range. In Fig.~\ref{fig:inty} we show the
integrated yields as a function of Glauber $N_\text{part}$. We find a power law
dependence with power $1.48\pm 0.08(\text{stat}) \pm 0.04(\text{syst})$ of the
integrated yield on $N_\text{part}$, independent of the considered \pt-range
reflecting the similar inverse slopes observed earlier. The yield of hadrons
grows roughly linear with $N_\text{part}$.





\bibliographystyle{elsarticle-num}







\end{document}